\documentclass[manuscript,screen]{acmart}
\AtBeginDocument{%
  }

\setcopyright{acmlicensed}
\copyrightyear{2018}
\acmYear{2018}
\acmDOI{XXXXXXX.XXXXXXX}
\acmConference[Conference acronym 'XX]{Make sure to enter the correct
  conference title from your rights confirmation email}{June 03--05,
  2018}{Woodstock, NY}
\acmISBN{978-1-4503-XXXX-X/2018/06}

\usepackage{listings}

\begin{document}

\title{Inferential Privacy Leakage in Anonymized Conversational AI Logs}

\author{S M Mehedi Zaman}
\author{Kiran Garimella}
\affiliation{%
  \institution{Rutgers University}
  \country{USA}
}








\renewcommand{\shortauthors}{Zaman and Garimella}

\begin{abstract}
  Conversations with ChatGPT have become a new kind of personal record: long, accumulated over months, and routinely touching health, work, family, and belief in ways search queries never did. These logs now reach many parties beyond the platform. They are shared with researchers, released as public datasets, retained under court order, and prepared for advertising, usually after names, places, and explicit identifiers have been removed, on the assumption that what remains is safe to hold and share. We test that assumption with complete ChatGPT histories donated, with consent, by 1,057 users in Brazil, India, Nigeria, and Pakistan, each paired with self-reported demographics, and for 212 of them with their Google Search and YouTube histories as well. Users volunteer personal information constantly: 34.5\% of 1.24 million messages contain it, and the median user does so within the first 14\% of their history. We then remove messages flagged for explicit demographic self-identification and ask an open-weights LLM to infer each user’s age bracket, gender, and country from the remaining history. It recovers country at weighted F1 0.86, gender at 0.76, and age bracket at 0.51, well above majority-class and feature-based baselines, with most of the gender and country signal already present in the first 5\% of a history. Redaction removes what users say about themselves but it does not remove what the model assumes about them. The model's rationales and its error structure show four stereotypes doing the work: technical content reads as male, technical skill reads as Western, fluent English reads as American, and contemporary interests read as young. In consequence, the model reads half of the women in the cohort as men, misreads roughly a quarter of Nigerian and half of Pakistani users, often as American or British, and correctly identifies no user aged 45 or above. The same users' search and watch histories show ChatGPT to be a comparable profiling surface, not a redundant one. The harms therefore split: users who match the model's defaults are exposed, users who do not are misattributed, and both groups sit outside the populations that prevailing privacy and fairness frameworks were built around. We draw out what this means for the inference criterion in anonymization standards, for research data releases, and for who bears the cost when a model trained on someone else's defaults fills in the blanks.
\end{abstract}




\maketitle

\section{Introduction}

Conversational AI platforms are rapidly transitioning from subscription-based productivity tools to ad-supported commercial ecosystems \citep{murgia2024openai_ads}. As operators introduce sponsored placements and behavioral targeting into conversational interfaces, the multi-turn interaction logs generated by millions of users have become high-value behavioral assets. Unlike disjointed search queries or clickstream logs \citep{yan2009much,mayer2012third,englehardt2016online}, conversational histories accumulate over months of continuous dialogue \citep{mireshghallah2024trust}. Users turn to conversational agents to debug code, seek medical explanations, draft sensitive workplace correspondence, and discuss religious practices. While these exchanges rarely contain explicit personal identifiers such as social security numbers or residential addresses, they record an extensive chronicle of a user's everyday life.

As conversational archives are retained under corporate terms of service, monetized for targeted advertising, or shared with academic researchers as public benchmark datasets \citep{chatterji2025people}, data custodians face an acute governance challenge. The prevailing regulatory and institutional response relies on standard text de-identification: redacting named entities, contact details, and explicit demographic declarations under the assumption that the remaining text is non-personal and safe to circulate (e.g., following HIPAA Safe Harbor or GDPR Recital 26 principles). This paper investigates whether that assumption holds, or whether modern large language models (LLMs) acting as inferential adversaries can accurately reconstruct sensitive demographic identities, such as age, gender, and nationality, from scrubbed conversational logs alone.

This challenge resists traditional privacy defenses because demographic identity in natural language is rarely confined to isolated tokens. Instead, identity is diffusely distributed across linguistic phrasing, syntactic style, topical combinations, and cultural framing. Prior work has demonstrated that data sanitization fails to guarantee natural language privacy because contextual correlations preserve sensitive signals \citep{brown2022does}, and that technical privacy regimes often prioritize input-data scrubbing while overlooking the sensitivity of inferred outputs \citep{chowdhary2025technical}. Because modern LLMs possess extensive world knowledge and statistical associations learned during pretraining \citep{staab2023beyond}, they can connect these subtle, dispersed markers across multiple conversation turns. Consequently, syntactic redaction provides a false sense of security: it excises what users explicitly state about themselves, while leaving intact the latent signals that allow automated systems to infer who they are.

This inferential vulnerability carries severe fairness and accountability risks, particularly when evaluated outside dominant Western contexts. Pretrained language models encode cultural, gender, and socio-economic heuristics inherited from predominantly Western pretraining corpora \citep{bender2021dangers,sheng2019woman}. When an open-weights model is tasked with demographic profiling on populations in the Global South, it does not function as an impartial statistical estimator. Instead, its inference engine operates through pretraining stereotypes. In algorithmic fairness, research from India and across the Global South warns against the uncritical transfer of Western assumptions and emphasizes how algorithmic harms emerge through local socio-cultural conditions and global development imbalances \citep{sambasivan2021reimagining,aneja2025beyond,mohamed2020decolonial}. In commercial deployment, such as an LLM-driven advertising network, this dynamic produces an asymmetric double harm. Users whose conversational behavior matches the model's dominant archetypes face rapid privacy exposure, with their latent demographic attributes unmasked from routine queries. Conversely, users whose identities deviate from Western and patriarchal norms, such as female software engineers, older technology users, or English-fluent professionals in Nigeria and Pakistan, face systematic misrepresentation and algorithmic erasure.

To evaluate these dynamics empirically, we study a multi-country dataset of complete ChatGPT conversation histories donated with informed consent by 1{,}057 users across four underrepresented Global South nations: Brazil, India, Nigeria, and Pakistan. Each conversational archive is paired with survey-verified ground truth for age bracket, gender, and country of residence. To establish how conversational profiling compares to legacy behavioral tracking, an expanded sub-cohort of 212 Indian participants also donated their matched Google Search, YouTube search, and YouTube watch histories alongside extended demographic attributes. We implement a multi-stage sanitization protocol that uses Named Entity Recognition (NER) to remove named entities and an LLM classifier to exclude messages containing explicit demographic self-declarations, ensuring that subsequent inference evaluations isolate the leakage arising purely from residual conversational style and topical mixture.

Our empirical findings demonstrate the fragility of current anonymization practices and the pervasive influence of model stereotypes. Auditing 1.24 million raw messages across a twenty-category taxonomy \citep{cogendez2026can}, we find that 34.5\% of user messages contain personal information, with the median user disclosing personal context within the first 14\% of their interaction. On the sanitized logs, an open-weights model (Llama-3.3-70B-Instruct) successfully reconstructs country of residence (weighted $F_1 = 0.86$), gender ($0.76$), and age bracket ($0.51$), with country and gender stably identified within the first 5\% of user history. However, an audit of 600 model-generated rationales shows that predictions rely heavily on four cultural stereotypes (technical topics read as male, technical skill as Western, fluent English as American, and modern interests as young). Consequently, the model misclassifies half of the women as men (female recall 0.50), misattributes roughly a quarter of Nigerian and half of Pakistani users as American or British, and correctly identifies no user over 45. Finally, our matched cross-platform comparison reveals that while search queries excel at concentrated, transactional signals like religion and income, conversational AI provides a distinct, narrative-driven substrate that excels at exposing traits like age, education, and voting preference.

This work makes four broad contributions: (i) we provide empirical evidence that standard text de-identification fails against LLM-driven inference, demonstrating to data custodians and repositories that scrubbing explicit PII does not render conversational archives non-personal; (ii) by showing how easily demographic traits are reconstructed from conversational context, we illustrate the privacy and fairness risks of emergent ad-targeting systems that operationalize automated behavioral profiling without user consent; (iii) we demonstrate that demographic inference in the Global South produces structural epistemic harms, where non-Western users are either surveilled when they fit Western defaults or systematically erased and misattributed when they do not; and (iv) using a matched multi-platform dataset, we provide the first direct empirical comparison between conversational AI logs and two decades of search and video tracking, charting how behavioral surveillance shifts as users move from transactional search to interactive dialogue.

\section{Related Work}
\label{sec:related_work}

Our study brings together three bodies of literature: privacy and inferential profiling in natural language, behavioral surveillance across digital substrates, and algorithmic stereotyping and cultural representation in large language models.

\subsection{Syntactic Sanitization vs.\ Inferential Privacy}

Data protection frameworks have traditionally focused on syntactic de-identification, which aims to prevent re-identification by scrubbing direct and indirect identifiers such as names, locations, and personal identification numbers (e.g., GDPR Recital 26 and HIPAA Safe Harbor). However, legal and technical scholarship increasingly demonstrates that syntactic sanitization does not prevent inferential profiling. \citet{wachter2019right} argue that privacy law overlooks the risks of inferred data, which allows automated systems to assemble sensitive behavioral and demographic profiles without holding explicit identity tokens. In natural language processing, \citet{brown2022does} show that text sanitization fails to guarantee privacy because contextual correlations and semantic associations preserve sensitive signals across documents. Similarly, \citet{chowdhary2025technical} find that technical privacy regimes often prioritize scrubbing input data while neglecting the sensitivity of attributes inferred by downstream models.

Empirical studies of human--LLM interactions confirm that users routinely disclose sensitive personal information. \citet{mireshghallah2024trust} document widespread disclosures across naturally occurring ChatGPT conversations, showing that standard Personally Identifiable Information (PII) detection fails to capture the full scope of personal context. \citet{cao2026dual} show that disclosure patterns are culturally differentiated, varying significantly across national contexts. To systematically categorize these disclosures, \citet{cogendez2026can} introduce a twenty-category personal-information taxonomy, which we adapt to message-level granularity in Section~\ref{sec:disclosure}. We build on this work by demonstrating that even after all explicit disclosures and named entities are eliminated, an LLM adversary can still infer sensitive demographic attributes from residual conversational style.

\subsection{Behavioral Profiling Across Digital Substrates}

Inferring demographic identity from online behavioral traces has formed the foundation of commercial digital advertising for more than two decades. Search queries and browsing histories have long been leveraged to infer gender, age, income, and commercial intent \citep{yan2009much}, supported by a pervasive infrastructure of third-party tracking cookies and behavioral monitoring \citep{mayer2012third,englehardt2016online}. In natural language processing, early work demonstrated that author age, gender, and personality traits could be recovered from public blogs and social media text using stylometric and lexical features \citep{schwartz2013personality,hovy2016social}. Recent work shows that pretrained LLMs can infer personal attributes from public Reddit profiles with high accuracy \citep{staab2023beyond}, and that models generate persistent demographic associations when prompted with personal names \citep{staufer2026llms}.

The shift from transactional search to conversational AI introduces a qualitatively different behavioral substrate. While search logs capture fragmented, intent-driven keyword queries, conversational logs capture continuous, multi-turn narratives where users reflect on professional dilemmas, personal relationships, and health concerns. Furthermore, the downstream consequences of demographic inference in these platforms are increasingly tangible. As commercial LLM providers deploy ad targeting and sponsored recommendations \citep{murgia2024openai_ads}, demographic inference becomes an operational profiling mechanism. Research on algorithmic advertising audits shows that demographic inference errors can distort ad delivery and conceal systemic disparities \citep{imana2025auditing}. Our matched cross-platform benchmark (Section~\ref{sec:crossplatform}) directly compares conversational logs with search and video watch histories for the exact same participants, establishing how identity signals diverge across these behavioral substrates.

\subsection{Algorithmic Stereotypes and Cultural Representation in the Global South}

Foundation models reproduce and amplify social stereotypes inherited from massive web pretraining corpora \citep{sheng2019woman,abid2021persistent}. As \citet{bender2021dangers} emphasize, scaling language models without addressing training data composition encodes dominant cultural viewpoints as universal defaults. In cross-cultural audits, models routinely project Western norms and social categories onto non-Western settings \citep{naous2024beer}.

These representational biases have severe implications when language models are deployed as profiling engines in the Global South. \citet{sambasivan2021reimagining} challenge the uncritical transfer of Western fairness metrics to India, arguing that algorithmic fairness must be grounded in localized socio-cultural realities. Similarly, \citet{aneja2025beyond} audit LLM gender bias in Indian contexts, finding that technical and professional cues are heavily gendered due to Western-centric data and local occupational structures. Connecting with broader calls for decolonial AI \citep{mohamed2020decolonial}, these studies demonstrate that demographic inference in non-Western populations cannot be treated as a neutral classification task. When a Western-trained model encounters users whose communication violates its normative defaults, such as women in software engineering or English-fluent professionals in Nigeria and Pakistan, the result is not random noise, but systematic misattribution and representational erasure.

\section{Dataset}
\label{sec:dataset}

We analyze two datasets collected under institutional IRB approval through a privacy-preserving data-donation pipeline recruiting consenting participants via Clickworker \cite{clickworker_platform}. Participants were informed of the research objectives, provided explicit consent to donate their exported interaction archives, and received financial compensation calibrated to exceed local hourly minimum wage benchmarks in each participating country. The first dataset is a multi-country corpus of donated ChatGPT conversation histories and baseline demographic surveys from four Global South countries. The second is an expanded sub-cohort of Indian participants who additionally donated their Google Search, YouTube search, and YouTube watch histories via Google Takeout alongside a detailed socio-demographic survey. Table~\ref{tab:datasets} summarizes both datasets. Data collection took place in February 2026, capturing users' complete conversation histories from account inception up to the collection date. Anonymized analysis scripts and schema examples are documented in Appendix~\ref{app:data_code}.

\begin{table}[ht]
\centering
\small
\caption{Overview of the two donated datasets used in this paper.}
\label{tab:datasets}
\begin{tabular}{@{}p{2.1cm}p{2.4cm}p{2.7cm}@{}}
\toprule
 & \textbf{Dataset~1} (multi-country) & \textbf{Dataset~2} (Indian sub-cohort) \\
\midrule
$N$ (analytic) & 1{,}057 & 212 \\
\addlinespace
Countries & Brazil, India, Nigeria, Pakistan & India only \\
\addlinespace
Data sources & ChatGPT conversation logs & ChatGPT logs, Google Search, YouTube Search, YouTube Watch \\
\addlinespace
Demographics & age bracket, gender, country & age, gender, country, religion, caste, education, income, voting preference \\
\bottomrule
\end{tabular}
\vspace{-\baselineskip}
\end{table}

\subsection{Multi-country ChatGPT corpus (Dataset~1)}
\label{sec:dataset_main}

Participants for Dataset~1 were recruited from four Global South countries historically underrepresented in privacy and language modeling audits: Brazil, India, Nigeria, and Pakistan. Each participant uploaded their complete, official ChatGPT conversation JSON export and completed a demographic survey recording their age (categorized into brackets: 18--24, 25--34, 35--44, and 45+), platform-recorded gender (male or female), and country of residence.

The raw donated corpus comprises 1{,}242{,}109 user messages. To isolate implicit demographic leakage from overt self-disclosure, we construct an analytic cohort using two filtering criteria: a conversation length floor excluding users in the bottom 10\textsuperscript{th} percentile ($\le 10$ user messages) to ensure sufficient context for evaluation, and a safety filter (Section~\ref{sec:clean_data}) using Llama-3.3-70B-Instruct to exclude any messages containing explicit demographic self-declarations. This yields a final analytic cohort of $N=1{,}057$ users: India ($n=456$), Nigeria ($n=206$), Brazil ($n=205$), and Pakistan ($n=190$).

We maintain a strict separation between audit and inference evaluations throughout the paper. The explicit disclosure audit (Section~\ref{sec:disclosure}) is computed across all 1{,}242{,}109 messages in the un-filtered corpus, providing an authentic baseline of real-world disclosure rates. Conversely, all demographic inference evaluations (Sections~\ref{sec:inference} and~\ref{sec:crossplatform}) are conducted exclusively on the 1{,}057-user analytic cohort, ensuring that the adversary cannot rely on direct self-declarations.

\subsection{Cross-platform Indian sub-cohort (Dataset~2)}
\label{sec:dataset_india}

Dataset~2 consists of an expanded sub-cohort of Indian participants ($N=212$) from Dataset~1 who consented to share their complete Google Takeout records, including Google Search query logs, YouTube search query logs, and YouTube watch histories, alongside an extended demographic questionnaire.

The expanded survey records five additional self-reported socio-political attributes: religion, caste category, monthly household income, highest educational attainment, and voting preference in the 2024 Indian general election. We evaluate inference performance across six attributes (age, gender, religion, education, income, and voting preference). Although caste data was collected to maintain consistency with standard Indian socio-demographic survey instruments, it was intentionally withheld from all inference experiments. Evaluating automated caste classification using Western-trained foundation models presents severe ethical risks, including the potential to legitimize algorithmic caste profiling and reinforce historical social stratifications \citep{sambasivan2021reimagining}. Within this 212-user sub-cohort, the age distribution comprises 18--24 ($n=82$), 25--34 ($n=81$), 35--44 ($n=31$), and 45+ ($n=18$); the gender breakdown is 174 men and 38 women.

\section{Methods}
\label{sec:methods}

Our measurement pipeline consists of three components: a disclosure audit (Section~\ref{sec:methods_audit}), a step-wise demographic-inference protocol (Section~\ref{sec:methods_inference}), and a qualitative analysis of natural-language reasoning rationales (Section~\ref{sec:methods_qual}). All components, as well as the analytic cohort filtering pipeline, use Llama-3.3-70B-Instruct in 4-bit quantization, selected over smaller open-weights candidates via human validation (Appendix~\ref{sec:methods_model}).

\subsection{Filtering and the analytic cohort}
\label{sec:clean_data}

We analyze user-authored messages exclusively, discarding model responses. We apply two filtering passes: a deterministic NER pass for disclosure auditing, and an LLM-based self-disclosure filter driving message exclusion.

\noindent\textbf{NER for audit.}
English user messages are processed through SpaCy's \texttt{en\_core\_web\_lg} pipeline to flag entities of type GPE, LOC, NORP, PERSON, ORG, and FAC. To prevent language-mismatch noise, NER is limited to English. The output of NER filtering is used in the disclosure audit (Section~\ref{sec:methods_audit}) but does not exclude users from the cohort.

\noindent\textbf{LLM-based self-disclosure filter.}
To identify explicit demographic statements that bypass entity recognition (such as ``As a single mother\ldots'' or ``Being a 22-year-old student\ldots''), we classify all user messages across all donor languages using Llama-3.3-70B-Instruct (prompt in Listing~\ref{lst:prompt}, Appendix~\ref{app:prompts}). Messages are labeled \emph{SAFE} or \emph{UNSAFE} based on direct demographic declarations regarding age, gender, familial role, religion, or ethnicity.

For each user, we retain messages labeled \emph{SAFE}. Joined with the conversation length floor ($>10$ messages; Section~\ref{sec:dataset_main}), this yields the final cohort of $N=1{,}057$ users. Consequently, subsequent inference evaluations (Sections~\ref{sec:inference} and~\ref{sec:crossplatform}) isolate demographic leakage originating from conversational style, topical preferences, and linguistic nuance alone.

\subsection{Disclosure audit}
\label{sec:methods_audit}

The disclosure audit reported in Section~\ref{sec:disclosure} is run on the full donated corpus of 1{,}242{,}109 user messages before analytic-cohort filtering, reflecting the broader donor population rather than the conservative subset used for inference.

\noindent\textit{Discovery point.}
For each user we identify the chronological index of the first message flagged as \emph{UNSAFE} and normalize by total user-message count:
\begin{equation}
\label{eq:pdiscovery}
    P_{\text{discovery}} = \frac{\text{Index}_{\text{first flag}}}{\text{Messages}_{\text{total}}} \times 100.
\end{equation}
Because total message counts vary across users, normalized $P_{\text{discovery}}$ serves as the comparable statistic across the cohort. Its distribution, mean, median, and turn-1 spike are reported in Section~\ref{sec:disclosure}.

\noindent\textit{Category classification.}
We classify each \emph{UNSAFE} message into twenty personal-information categories following \citet{cogendez2026can}. We adapt their taxonomy in two ways: applying it at message-level granularity (allowing a conversation to span multiple categories) and validating taxonomy fit by manually coding 200 sampled \emph{UNSAFE} messages, which produced no categories outside the original twenty. The remaining flagged messages are labeled with Llama-3.3-70B-Instruct using Listing~\ref{lst:classifier_prompt} (Appendix~\ref{app:prompts}).

\noindent\textit{Qualitative analysis.}
Alongside checking taxonomy fit, we qualitatively examined stratified sample of 200 messages to identify
illustrative examples of personal disclosures and category
boundaries. Among those, we include ten examples in
Appendix~\ref{app:disclosure_examples}.


\noindent\textit{Disclosure accumulation.}
For each participant, we calculate the cumulative fraction
of their recorded flagged messages encountered at each normalized
history position, then average these fractions with equal
weight per participant. We estimate 95\% pointwise percentile
confidence intervals using 5{,}000 bootstrap resamples of
participants, retaining each participant's complete trajectory
within each resample.

\subsection{Step-wise demographic inference}
\label{sec:methods_inference}

The inference protocol underlying Sections~\ref{sec:inference} and~\ref{sec:crossplatform} predicts each user's demographic attributes from sanitized conversation history alone, using progressively larger prefixes of that history.

\noindent\textbf{Tasks and ground truth.}
For each analytic-cohort user, Llama-3.3-70B-Instruct predicts age bracket (four categories: 18--24, 25--34, 35--44, 45+), gender (binary, from donation platform records), and country of residence (open-ended generation). Ground truth is obtained from the initial demographic survey (Section~\ref{sec:dataset_main}). Demographics prompts (Listings~\ref{lst:country_prompt}, \ref{lst:gender_prompt}, and \ref{lst:age_prompt} in Appendix~\ref{app:prompts}) employ a chain-of-thought structure producing a 2--3 sentence rationale before the final label, which boosts prediction accuracy (Appendix~\ref{sec:methods_model}) and provides interpretability for qualitative analysis (Section~\ref{sec:methods_qual}).

\noindent\textbf{Incremental-prefix protocol.}
For every (user, attribute) pair, we run the prompt twenty times across chronological message history prefixes of size $k \in \{5\%, 10\%, 15\%, \ldots, 100\%\}$. At each prefix, the model returns a rationale and label, totaling $\approx 63{,}000$ prompted inferences across three attributes for $N=1{,}057$ users.

\noindent\textbf{Outcome metrics.}
We evaluate predictions across all prefix thresholds $k \in \{5\%, 10\%, \ldots, 100\%\}$. We define the \emph{context-needed} threshold $k^*$ as the smallest prefix where prediction matches ground truth and remains stable for all subsequent prefixes $k' \ge k^*$ (undefined if stability is never achieved up to $100\%$). Primary metrics (F1, precision, recall in Section~\ref{sec:inference}) are evaluated at the final $100\%$ prefix. Qualitative analysis (Section~\ref{sec:methods_qual}) examines rationales generated at $k^*$ for stable successes, and at $100\%$ otherwise.

Although exact ages are available in the demographic survey,
we evaluate age inference as four-bracket classification
(18--24, 25--34, 35--44, and 45+). We therefore report
classification metrics rather than years-based estimation
error, without converting predicted brackets into point ages.

\noindent\textbf{Cross-platform replication.}
For the 212 users in Dataset~2 (Section~\ref{sec:dataset_india}), we apply this incremental-prefix protocol to three additional data streams replacing ChatGPT history: Google Search query logs, YouTube search query logs, and YouTube watch histories. Demographic prompts remain identical except for swapping the data-source descriptor. We additionally predict religion, education level, monthly income, and voting preference using matching survey options, while dropping country (as all 212 users are Indian). Exemplar Google-search prompts (Listings~\ref{lst:age_prompt_google}--\ref{lst:vote_prompt_google}) are in Appendix~\ref{app:prompts}.

\noindent\textbf{Methodological parity between filter and adversary.}
Using Llama-3.3-70B-Instruct as both safety filter (Section~\ref{sec:clean_data}) and demographic adversary ensures the adversary holds no informational advantage over the filter. Any inferred demographic signal originates from conversational content that was previously evaluated and certified as non-disclosive by the exact same model architecture. While we do not claim a universal lower bound, as proprietary frontier models might recover even more signal, this parity controls for filter-adversary mismatch under a transparent open-weights setup.

\subsection{Qualitative Analysis of Model Rationales}
\label{sec:methods_qual}

Natural-language rationales generated at context-needed prefixes $k^*$ serve as empirical input for analyzing the model's reasoning heuristics and latent cultural assumptions (Sections~\ref{sec:inference} and~\ref{sec:crossplatform}).

\noindent\textbf{Sampling.}
For Dataset~1, we drew a class-stratified random sample of 200 rationales per attribute ($600$ total across age, gender, country) to ensure adequate representation of minority classes. For Dataset~2, we analyzed all 212 users across all attribute-platform combinations without sub-sampling.

\noindent\textbf{Coding and validation.}
We conducted inductive thematic coding on sampled rationales to derive the primary stereotype tropes detailed in Section~\ref{sec:inference} (\emph{Tech~$\equiv$~male}, \emph{Tech~$\equiv$~Western}, \emph{English fluency~$\equiv$~US}, and \emph{Contemporary content~$\equiv$~young}). To confirm that these patterns generalize across the full corpus, we ran a scaled lexical validation using diagnostic term clusters, confirming that sampled qualitative patterns are broadly distributed across all generated rationales. The complete coding protocol, inter-coder reliability metrics, and diagnostic dictionaries appear in Appendix~\ref{app:qual_methodology}.

\section{Results}

Our evaluation consists of three parts: auditing explicit user self-disclosures in raw conversation histories (Section~\ref{sec:disclosure}); evaluating implicit demographic recovery on an anonymized cohort (Section~\ref{sec:inference}); and benchmarking ChatGPT against search and watch histories (Section~\ref{sec:crossplatform}).

\subsection{What users disclose explicitly}
\label{sec:disclosure}

Evaluating all 1{,}242{,}109 user messages in the raw corpus against the twenty disclosure categories of \citet{cogendez2026can} via Llama-3.3-70B-Instruct, we find that 428{,}865 messages (34.5\%) contain personal information.

Table~\ref{tab:data_sharing} summarizes the category distribution. Three categories comprise over half of all flagged messages: \emph{job and education} (25.1\%), \emph{lifestyle and habits} (18.7\%), and \emph{mental state} (11.6\%). Explicit demographic statements are rare (age: 1.5\%, gender: 0.3\%, sexual orientation: 0.1\%), showing that users primarily reveal identity through contextual disclosures rather than direct assertions.

\begin{table}[ht]
\centering
\small
\caption{Distribution of flagged messages across the twenty disclosure categories of \citet{cogendez2026can} ($N=428{,}865$ flagged messages out of 1{,}242{,}109 user messages).}
\label{tab:data_sharing}
\begin{tabular}{lr}
\toprule
\textbf{Category} & \textbf{\% of flagged} \\
\midrule
Job and education                   & 25.14 \\
Lifestyle and habits                & 18.70 \\
Mental state, personality, mood     & 11.60 \\
Location and mobility               & 10.97 \\
Wealth, salary                      & 5.88  \\
Family life and relationships       & 3.95  \\
Physical health, diagnosis          & 3.87  \\
Religion                            & 2.98  \\
Ethnicity and citizenship           & 2.40  \\
Physical traits                     & 2.01  \\
Personal identifiers                & 2.00  \\
Account credentials                 & 1.87  \\
Recreational consumption            & 1.78  \\
Sexual and dating activities        & 1.71  \\
Political views                     & 1.50  \\
Age                                 & 1.47  \\
Mental health                       & 1.42  \\
Criminal records                    & 0.37  \\
Gender                              & 0.27  \\
Sexual orientation                  & 0.11  \\
\bottomrule
\end{tabular}
\end{table}

Country-level breakdowns (Appendix~\ref{app:country_breakdown}, Table~\ref{tab:country_distribution}) reveal distinct category patterns: Brazilian users over-index on Mental State (16.9\% vs.\ 5--10\% elsewhere); Nigerian users on Location (19.1\%) and Ethnicity (5.2\%); and Pakistani users on Job/Education (31.8\%) and Wealth (8.2\%). Indian users align closely with cohort averages.

\noindent\textbf{How personal information appears in messages.}
Our qualitative examination of the 200-message sample
illustrates how personal information appears within everyday
requests for help, including discussions of education,
finances, relationships, and health. The examples also
highlight category boundaries: descriptions of fear or
detachment do not by themselves establish a
diagnosis, while first-person statements in riddles do not
constitute personal disclosure. Examples are shown
in Appendix~\ref{app:disclosure_examples}.

\begin{figure}[htbp]
    \centering
    \includegraphics[width=0.75\linewidth]{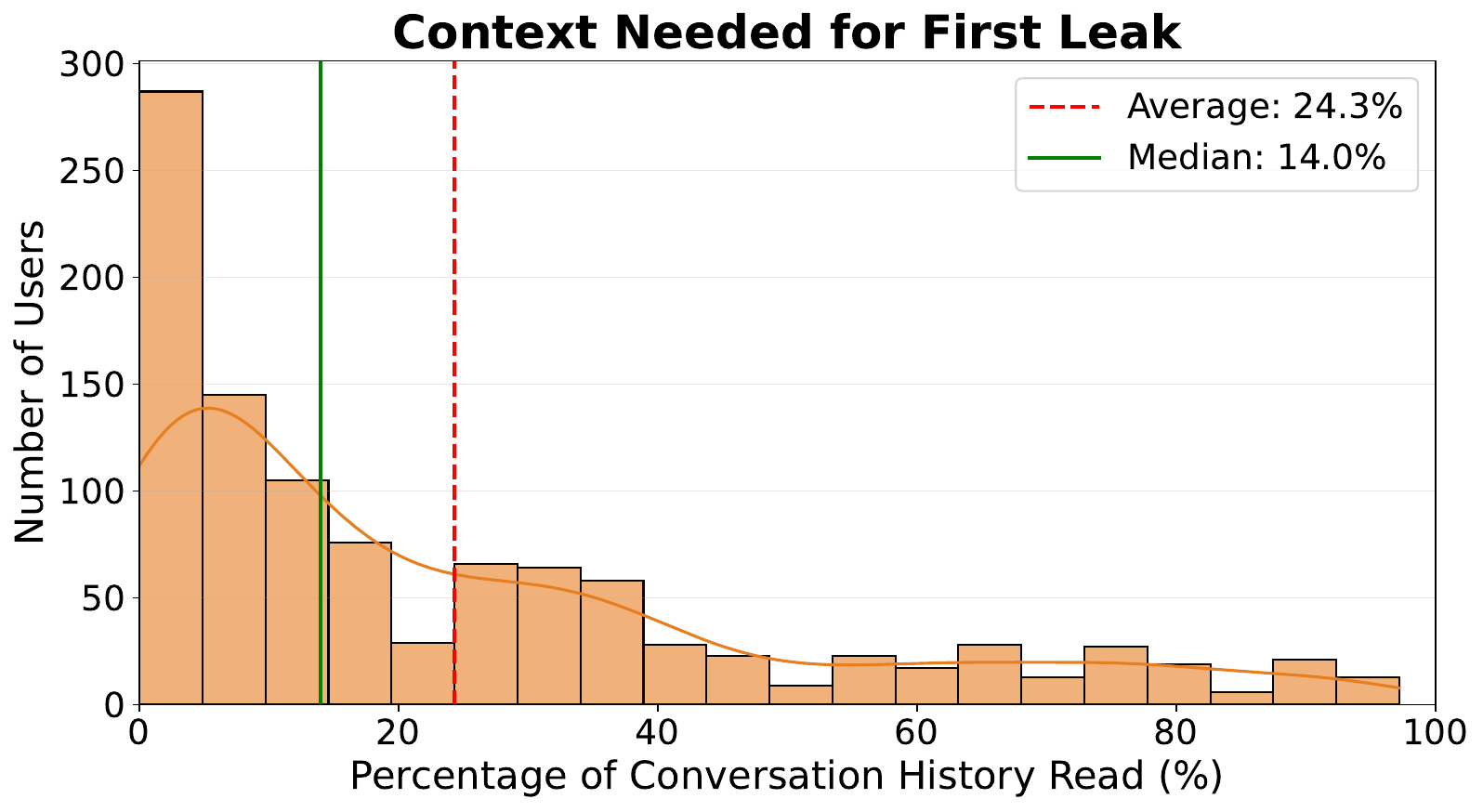}
    \caption{Distribution of discovery points $P_{\text{discovery}}$, representing the fraction of history preceding a user's first flagged disclosure.}
    \label{fig:privacy_context}
\end{figure}

\noindent\textbf{Discovery point distribution.}
Normalizing the message index of a user's first disclosure by their total message count ($P_{\text{discovery}}$, Eq.~\ref{eq:pdiscovery}) yields a median of 14.0\% and mean of 24.3\% (Figure~\ref{fig:privacy_context}). A distinct spike at $P_{\text{discovery}}=0$ highlights users who disclose personal content in their initial interactions.

\begin{figure}[htbp]
    \centering
    \includegraphics[width=0.85\linewidth]
    {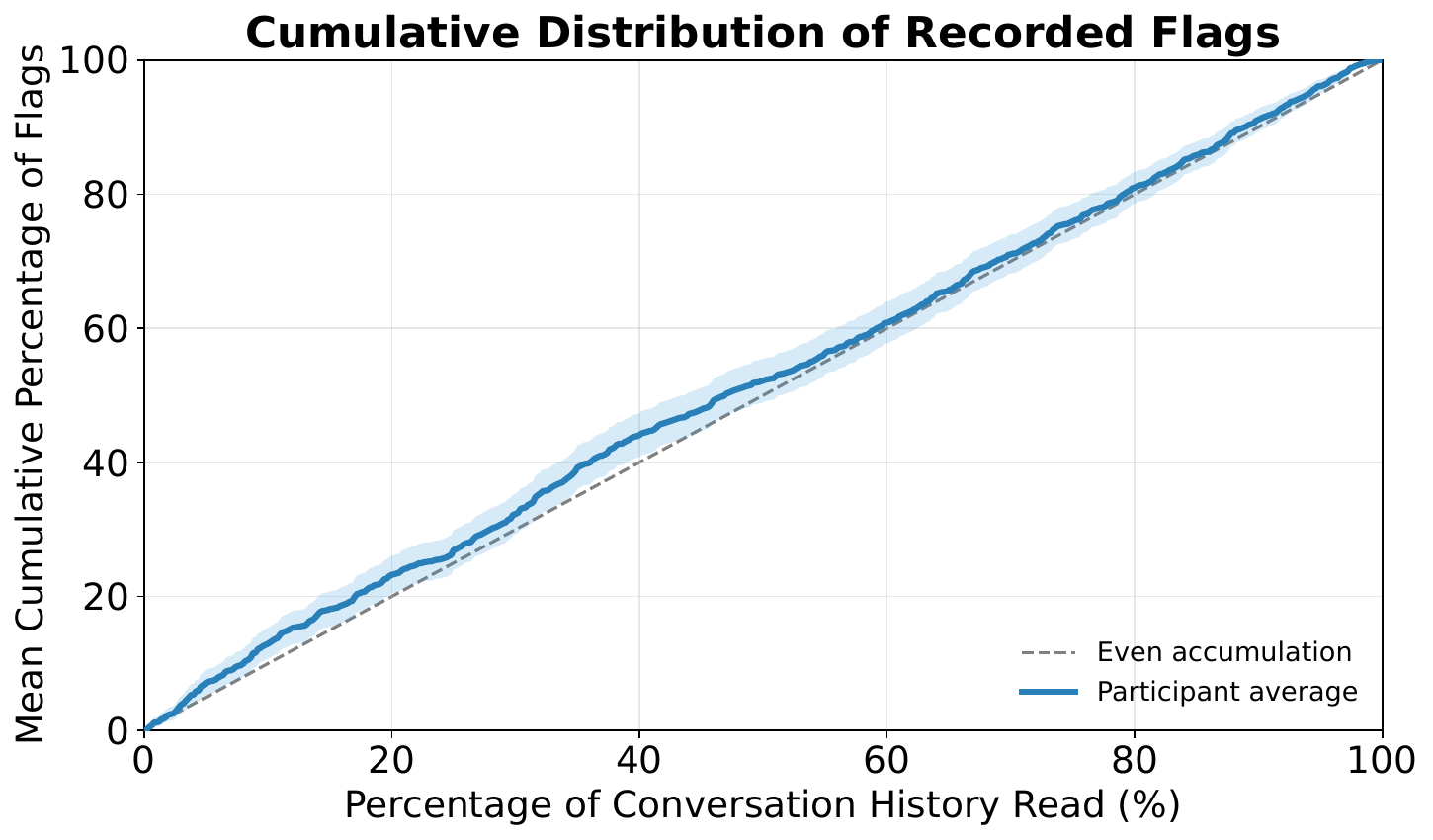}
    \caption{Participant-averaged cumulative distribution
    of recorded flag positions across observed message
    histories.
    Each participant receives equal weight.
    Shading indicates 95\% pointwise participant-bootstrap
    confidence intervals. The dashed diagonal represents
    even accumulation.}
    \label{fig:leak_cdf}
\end{figure}

\noindent\textbf{Disclosure throughout conversation histories.}
Figure~\ref{fig:leak_cdf} shows the cumulative percentage
of each participant's flagged messages encountered as more
of their history is read, averaged equally across all participants. Flags occur throughout the observed histories,
with some concentration earlier: the first half contains,
on average, 52.2\% of a participant's recorded flags. Personal information therefore continues to appear in later interactions rather than being confined to a user's initial conversations.

The remaining evaluations analyze the conservative $N=1{,}057$ analytic cohort, isolating what an adversary recovers when all explicit disclosures are filtered out.

\subsection{Demographic inference from sanitized logs}
\label{sec:inference}

We evaluate how much demographic identity an open-weights LLM recovers from conversation histories stripped of all explicit self-disclosures. Following Section~\ref{sec:methods_inference}, Llama-3.3-70B-Instruct evaluates progressively larger chronological message prefixes ($5\%$ increments). Classification metrics are evaluated at the final 100\% prefix; $k^*$ is used separately to summarize the context required for stable correctness.

\noindent\textbf{Identity recovery exceeds baselines.}
Table~\ref{tab:inference_per_class} compares per-class performance against a majority-class baseline and an XGBoost feature-based baseline (Appendix~\ref{app:xgb_methodology}). The LLM achieves weighted $F_1$ scores of $0.51$ for age, $0.76$ for gender, and $0.86$ for country, substantially outperforming majority-class ($0.23$, $0.52$, $0.26$) and XGBoost ($0.48$, $0.64$, $0.52$) baselines. Identifying Brazil ($F_1 = 0.89$) and India ($0.91$) from scrubbed text is particularly notable.

\begin{table}[ht]
\centering
\small
\caption{Per-class inference performance across demographic
attributes. Block headers report overall weighted $F_1$
for the Majority, XGBoost, and LLM models.}
\label{tab:inference_per_class}

\begin{tabular}{@{}lcccr@{}}
\toprule
& Prec. & Recall & $F_1$ & Support \\
\midrule
\multicolumn{5}{@{}l@{}}{\emph{Age bracket}
(Maj. $F_1=0.23$, XGB $=0.48$, LLM $=0.51$)} \\
18--24 & 0.60 & 0.69 & 0.64 & 389 \\
25--34 & 0.53 & 0.51 & 0.52 & 424 \\
35--44 & 0.36 & 0.35 & 0.36 & 185 \\
45+    & 0.00 & 0.00 & 0.00 & 59  \\
\midrule
\multicolumn{5}{@{}l@{}}{\emph{Gender}
(Maj. $F_1=0.52$, XGB $=0.64$, LLM $=0.76$)} \\
Female & 0.76 & 0.50 & 0.61 & 360 \\
Male   & 0.79 & 0.89 & 0.84 & 697 \\
\midrule
\multicolumn{5}{@{}l@{}}{\emph{Country of residence}
(Maj. $F_1=0.26$, XGB $=0.52$, LLM $=0.86$)} \\
Brazil   & 0.98 & 0.81 & 0.89 & 205 \\
India    & 0.88 & 0.95 & 0.91 & 456 \\
Nigeria  & 0.99 & 0.75 & 0.85 & 206 \\
Pakistan & 0.98 & 0.53 & 0.69 & 190 \\
\bottomrule
\end{tabular}%

\end{table}

\noindent\textbf{Rapid identity leakage.}
Gender and country unmask rapidly: both achieve a median context-needed threshold at our $5\%$ protocol floor (Table~\ref{tab:context_needed}; full performance trajectories across 5\%--100\% prefixes and sub-group breakdowns at key checkpoints appear in Appendix~\ref{app:prefix_performance}, Tables~\ref{tab:country_prefix_performance}--\ref{tab:age_breakdown}). Conversely, age classification requires substantially longer interaction histories (median $70\%$, mean $53.8\%$). Gender unmasks faster for men (mean $21.1\%$) than women ($36.7\%$).

\begin{table}[ht]
\centering
\small
\caption{Conversation context (\% of total messages) required for stable correct prediction. Medians at 5\% indicate the protocol floor is reached.}
\label{tab:context_needed}
\begin{tabular}{@{}lcc@{}}
\toprule
Attribute / group & Mean & Median \\
\midrule
Age                   & 53.8\% & 70\% \\
Gender (overall) & 24.6\% & 5\% \\
\quad Female     & 36.7\% & 20\% \\
\quad Male       & 21.1\% & 5\% \\
Country (overall)     & 19.8\% & 5\% \\
\quad Brazil          & 22.2\% & 5\% \\
\quad India           & 17.6\% & 5\% \\
\quad Nigeria         & 20.0\% & 5\% \\
\quad Pakistan        & 25.1\% & 10\% \\
\bottomrule
\end{tabular}
\vspace{-\baselineskip}
\end{table}

\noindent\textbf{Reasoning mechanisms: four recurring patterns.}
Analyzing rationales across 600 sampled predictions \citep{turpin2023language} reveals four dominant stereotype-mediated patterns driving classification (Appendix~\ref{app:bias_patterns}, Table~\ref{tab:bias_patterns}): (i) \emph{Tech $\equiv$ male}, where technical queries concerning Linux, software engineering, and finance systematically trigger male predictions, leading the model to misclassify 165 of 360 women as men (pulling female recall down to 0.50, compared to 0.89 for men); (ii) \emph{Tech $\equiv$ Western}, where rationales for Nigerian and Pakistani tech professionals systematically cite advanced technical proficiency to predict US or UK residency, driving Pakistani recall down to 0.53; (iii) \emph{English fluency $\equiv$ US}, where fluent English lacking distinct local slang or currency defaults to US residency, whereas Brazil achieves high recall (0.81, $F_1=0.89$) because Portuguese serves as an un-overridable geographic anchor; and (iv) \emph{Contemporary content $\equiv$ young}, where tech-literate older adults (such as users over 45 discussing 3D modeling software) are demoted to younger brackets, causing recall in the 45+ bracket to collapse to 0.00.

A prediction run provides additional
evidence of this geographic prediction pattern: 113 of
124 valid out-of-cohort country predictions (91.1\%)
name the United States or United Kingdom
(Appendix~\ref{app:out_of_cohort_countries},
Table~\ref{tab:out_of_cohort_countries}).

These results show that demographic inference relies on two mechanisms: direct residual markers and stereotype-mediated priors. Errors concentrate disproportionately on non-stereotypical groups, specifically women in technical roles, Global South developers, and older technology users.

\subsection{ChatGPT versus search and watch history as inference surfaces}
\label{sec:crossplatform}

To evaluate whether ChatGPT logs offer a novel inference surface or mirror web behavior, we apply our incremental-prefix protocol across four data streams (ChatGPT, Google Search, YouTube Search, YouTube Watch) for the $N=212$ Indian sub-cohort across six attributes.

\noindent\textbf{Complementary platform surfaces.}
Table~\ref{tab:demographic_comparison} reports weighted $F_1$ across platforms. No single stream dominates. ChatGPT logs provide the strongest signal for age ($F_1 = 0.54$), education level ($0.78$), and voting preference ($0.52$), where identity markers are embedded in rich context. Search logs outperform on gender (Google Search $0.90$), religion (YouTube Search $0.83$), and income (Google Search $0.62$), attributes surfacing through focused queries (detailed per-platform per-class breakdowns and qualitative case studies appear in Appendix~\ref{app:per_platform}). YouTube Watch has the lowest weighted \(F_1\) on five of six attributes due to passive consumption noise.

\begin{table*}[htbp]
\small
\centering
\caption{Cross-platform weighted $F_1$ for the $N=212$ Indian sub-cohort (detailed per-platform per-class metrics and qualitative case studies appear in Appendix~\ref{app:per_platform}). Bold = best per row.}
\label{tab:demographic_comparison}
\begin{tabular}{@{}lcccc@{}}
\toprule
\textbf{Attribute} & \textbf{ChatGPT logs} & \textbf{Google Search} & \textbf{YouTube Search} & \textbf{YouTube Watch} \\ \midrule
Age bracket         & \textbf{0.54}          & 0.44                   & 0.49                    & 0.39                   \\
Gender              & 0.88                   & \textbf{0.90}          & 0.89                    & 0.86                   \\
Religion            & 0.71                   & 0.75                   & \textbf{0.83}           & 0.72                   \\
Education level     & \textbf{0.78}          & 0.73                   & 0.68                    & 0.67                   \\
Monthly income      & 0.56                   & \textbf{0.62}          & 0.55                    & 0.51                   \\
Voting preference   & \textbf{0.52}          & 0.46                   & 0.43                    & 0.29                   \\
\bottomrule
\end{tabular}
\vspace{-\baselineskip}
\end{table*}

\noindent\textbf{Income and voting remain difficult.}
Across all surfaces, monthly income (best $F_1 = 0.62$) and voting preference (best $F_1 = 0.52$) are hardest to recover.

\noindent\textbf{Efficiency of search vs.\ conversation.}
Evaluating average context-needed ($k^*$), gender unmasks fastest on YouTube Search ($7.1\%$), YouTube Watch ($8.9\%$), and Google Search ($10.3\%$), compared to ChatGPT ($12.0\%$). For age, required context scales from YouTube Search ($9.0\%$) and Google ($12.1\%$) to ChatGPT ($15.7\%$) and YouTube Watch ($16.1\%$). Search queries offer concentrated, intent-dense signals, whereas ChatGPT logs dilute markers across broader topics.

\noindent\textbf{Implication.}
ChatGPT represents a distinct profiling surface comparable to traditional web search and video tracking. While older surfaces excel at transactional attributes (gender, religion, income), ChatGPT logs expose narrative-driven traits (age, education, voting). Effective anonymization must account for cross-platform leakage spanning these complementary behavioral substrates.

\section{Discussion}
\label{sec:discussion}

\subsection{The Illusion of `Anonymized' Conversational Data}
\label{sec:disc_findings}

Our findings demonstrate that conventional approaches to data sanitization fail when applied to conversational AI. In both regulatory doctrine (such as GDPR Recital 26 and HIPAA Safe Harbor) and open-science research (such as public releases of chat benchmarks and interaction corpora \citep{chatterji2025people}), data custodians routinely treat text as de-identified once named entities, contact information, and explicit self-declarations have been excised. Our results show that this syntactic definition of anonymization is an illusion. Even after removing 100\% of explicit demographic self-declarations and applying an NER filter, an open-weights LLM recovers country of residence at a weighted $F_1$ of $0.86$ and gender at $0.76$, with stable identification occurring within the first 5\% of a user's messages.

This empirical reality confirms the theoretical arguments of \citet{brown2022does}: natural language privacy cannot be guaranteed by token removal because semantic, syntactic, and topical correlations inherently encode identity. A single message asking for debugging assistance or advice on a dispute may appear benign in isolation, but across an extended multi-turn history, these diffuse signals aggregate into an unmistakable demographic footprint. For data repositories, open-science platforms, and institutional review boards, these findings demand a fundamental shift in governance. Data custodians cannot assume that releasing scrubbed conversational records is harmless. Sharing conversational data requires governance mechanisms that extend beyond syntactic filtering, including controlled-access enclaves, binding data-use agreements that prohibit re-identification and profiling, and continuous post-filtering inferential audits.

\subsection{Inferential Profiling and the Political Economy of Conversational Ads}
\label{sec:disc_ads}

The emergence of commercial advertising in conversational AI platforms marks a significant evolution in digital surveillance \citep{murgia2024openai_ads}. For two decades, behavioral targeting relied on tracking cookies, search queries, and clickstream monitoring to infer consumer intent and demographic profiles \citep{yan2009much,mayer2012third}. In conversational AI, platforms no longer require third-party tracking infrastructure. The conversational interface itself acts as an inferential engine, capable of assembling rich demographic and psychographic profiles directly from routine user prompts. 

This mechanism exposes a critical gap in contemporary privacy law. As \citet{wachter2019right} argue, data protection regimes like the GDPR grant individuals rights over their collected personal data (input data), but offer virtually no protection against inferred data produced through algorithmic reasoning. When an LLM infers that a user asking about software development is male, or that an English-fluent user in Nigeria is in the United States, that inference is generated dynamically without requiring the user to disclose demographic tokens. As conversational systems are integrated into advertising exchanges, these inferences can be operationalized to allocate high-value credit, employment, or housing opportunities without user awareness or consent. Furthermore, as \citet{imana2025auditing} demonstrate in traditional ad audits, demographic inference errors can conceal discriminatory delivery patterns from external oversight. Without regulatory standards that restrict the generation, retention, and commercialization of inferred demographic profiles, conversational AI risks accelerating behavioral surveillance under the guise of an interactive assistant.

\subsection{The Asymmetric Double Harm: Exposure vs.\ Epistemic Erasure}
\label{sec:disc_equity}

Our error analysis and qualitative coding reveal that the harms of demographic inference are distributed asymmetrically across populations. Rather than operating as an objective statistical classifier, the model relies on pervasive cultural, gender, and geographic heuristics learned during pretraining. These heuristics generate two distinct, diametrically opposed harms.

First, users whose communication patterns conform to the model's dominant archetypes experience severe privacy exposure. For male software developers or younger technology enthusiasts, identity is unmasked almost instantaneously, rendering syntactic anonymization ineffective. Second, users whose identities deviate from Western and patriarchal baselines suffer systematic misclassification and epistemic erasure. In our cohort, the model misclassifies 50\% of women as men, misattributes roughly a quarter of Nigerian and half of Pakistani users as American or British, and correctly identifies zero participants over age 45. In each case, the model's rationales explicitly cite professional or technical competence as justification for pushing predictions toward Western, male, or youthful defaults.

These disparate failure modes demonstrate that classification error in the Global South cannot be dismissed as harmless predictive noise. When demographic inferences are used to personalize user experiences, moderate content, or deliver targeted opportunities, these systematic misclassifications compound existing societal inequities \citep{calvi2024unfair}. For minoritized groups, such as female engineers in India or tech professionals in West Africa, the system commits a form of representational violence: users must either accept invasive surveillance if their identity is inferred, or endure algorithmic misgendering, cultural erasure, and economic misallocation if it is not.

\subsection{Reflexivity, Ethics, and Governance in the Global South}
\label{sec:disc_governance}

Interpreting these findings requires reflexivity regarding our methodology and the socio-cultural realities of our cohort. While our four-country dataset (Brazil, India, Nigeria, and Pakistan) broadens the empirical base of privacy research, our participants recruited through Clickworker skew younger, more tech-literate, and more fluent in English than their respective national populations. This selection reflects the contemporary user base of early conversational AI adopters in these regions, but it does not represent the Global South in its full socio-economic diversity.

Furthermore, we must critically evaluate the demographic categories themselves. Ground truth for gender was constrained to a binary format (male and female) by the demographic reporting structure of the crowdworking platform, a structural limitation that does not capture non-binary and gender-diverse identities. Similarly, while we collected self-reported caste data for the Indian sub-cohort, we made the deliberate ethical decision to withhold caste from our inference experiments. Predicting caste categories through foreign-trained foundation models risks legitimizing digital caste profiling and reinforcing historical caste hierarchies \citep{sambasivan2021reimagining}.

These ethical considerations reinforce the calls of \citet{sambasivan2021reimagining}, \citet{aneja2025beyond}, and \citet{mohamed2020decolonial} for decolonial approaches to algorithmic governance. Applying Western fairness metrics or technical privacy solutions to Global South contexts overlooks the historical and structural power imbalances embedded in global AI infrastructure. Privacy and fairness frameworks cannot be treated as universal plug-and-play standards; they must be grounded in the local socio-technical realities of the communities they govern.

\section{Limitations.}
Our evaluation is subject to three constraints: (1)~\emph{Adversary scope}: we test Llama-3.3-70B; proprietary models may exhibit higher inference capability due to compute scaling or lower capability due to safety refusals; (2)~\emph{Cohort selection}: our Clickworker cohort skews young, tech-literate, and English-fluent, which may facilitate easier demographic recovery than general populations; and (3)~\emph{Downstream scope}: we measure inference potential directly rather than downstream commercial ad-targeting pipelines.

\section{Conclusion}
Sanitized ChatGPT histories retain demographic signals which an off-the-shelf LLM can use to predict
age bracket, gender, and country of residence with weighted $F_1$ scores of $0.51$, $0.76$, and $0.86$,
respectively. Analysis of model-generated explanations identifies
recurring stereotyped associations alongside uneven prediction
errors across demographic groups. Comparisons with the same
participants' search and watch histories position ChatGPT as a
comparable surface for demographic profiling, with relative
performance varying by attribute and platform. These findings
show that message-level filtering alone does not prevent
demographic inference and motivate privacy evaluations that
consider both exposure through correct predictions and
misrepresentation through incorrect ones.

\clearpage

\section*{Endmatter}
\subsection*{Generative AI Usage Statement}
The initial draft of this paper was written by the authors without any AI assistance. Later, we used GPT-6.1 Sol to rephrase, shorten parts, and organize the manuscript. We did not use any generative AI tool to create tables, figures, or perform data analysis. However, we used the same GPT model to format the whole manuscript. We verified and are fully responsible
for all resulting text, code, calculations, references, and interpretations.

\subsection*{Ethical considerations}
This study was conducted under approval from our institutional IRB. All data were donated through Clickworker after explicit informed consent at recruitment, with participants told that their ChatGPT conversation histories, Google and YouTube exports (for the cross-platform sub-cohort), and demographic-survey responses would be used for research on what AI systems can infer about users. Compensation followed Clickworker's standard rates. We worked only on PII-stripped versions of the data for analysis, and restricted access to the immediate research team. Caste data was collected from the Indian sub-cohort but is held out of the present analysis given the particular sensitivity of caste-based prediction by a foreign-developed LLM in the Indian context \citep{sambasivan2021reimagining}; that analysis is reserved for a separate study with additional safeguards. We do not share inferences about individual donors outside the research team, and we do not measure downstream commercial or institutional consequences of the inferences we demonstrate.

\subsection*{Adverse impact}
This paper documents an inference-attack capability that is available today to anyone with a GPU and a publicly released open-weights LLM. We do not introduce a novel attack technique; we measure the consequences of running a routine one against real donated data. We acknowledge that publishing demonstrations of inference attacks can, in principle, enable the attacks they describe. In our judgment, the public-interest value of characterizing the capability, its asymmetric error distribution, and the populations it most exposes outweighs the marginal contribution to attack capacity that this paper makes. We have not released and will not release the underlying conversation data. The bias-pattern analysis in Section~\ref{sec:inference} names specific groups who are most exposed to misclassification; we are aware that naming these populations may also make them legible to actors who would target them, and we have judged the equity value of making the differential exposure visible to outweigh that risk.

\bibliographystyle{ACM-Reference-Format}
\bibliography{sample-base}

\appendix
\section{Appendix}

\subsection{Data and Code Availability}
\label{app:data_code}

Our anonymized codebase and replication materials, including mock conversation export schemas to preserve donor confidentiality, are openly accessible at: \url{https://anonymous.4open.science/r/inferential-privacy-leakage-2BB1}.

\subsection{Model selection and validation}
\label{sec:methods_model}

For each of the three LLM-driven classification tasks in the paper (SAFE/UNSAFE filtering, twenty-category disclosure classification, and demographic prediction) we evaluated four open-weights candidates at 4-bit quantization: Llama-3.1-8B, Mistral-Nemo-12B, Qwen3-32B, and Llama-3.3-70B-Instruct. For each (task, candidate) pair, one author manually verified outputs on 100 randomly sampled inputs against ground truth: against the Clickworker survey for demographic prediction, and against the author's own labels for the two classifier tasks (where no external ground truth exists). Llama-3.3-70B-Instruct outperformed the smaller candidates on every task. Table~\ref{tab:human_eval} reports the demographic-prediction accuracies on the 100-sample evaluation set; the gap to the next-best model is five to six percentage points across attributes. We adopt the 4-bit quantized Llama-3.3-70B-Instruct for all main analyses.

\begin{table}[ht]
\centering
\small
\caption{Accuracy on 100 randomly sampled demographic predictions per attribute, evaluated against Clickworker ground truth. Llama-3.3-70B-Instruct (4-bit) is used throughout the rest of the paper. Bold = best performance.}
\label{tab:human_eval}
\begin{tabular}{@{}lcccc@{}}
\toprule
\textbf{Model} & \textbf{Params} & \textbf{Country} & \textbf{Gender} & \textbf{Age} \\ \midrule
Llama-3.1         & 8B             & 0.68            & 0.62            & 0.38         \\
Mistral-Nemo     & 12B            & 0.72            & 0.66            & 0.41         \\
Qwen3          & 32B            & 0.77            & 0.71            & 0.46         \\
\textbf{Llama-3.3} & \textbf{70B}   & \textbf{0.83}   & \textbf{0.76}   & \textbf{0.51} \\ \bottomrule
\end{tabular}
\end{table}

\subsection{Prompts}
\label{app:prompts}

This appendix reproduces every prompt referenced in the paper, in three groups: the two prompts used to construct the analytic cohort and run the disclosure audit (Sections~\ref{sec:clean_data}, and~\ref{sec:methods_audit}); the three demographic-prediction prompts applied to ChatGPT conversations (Section~\ref{sec:methods_inference}); and the cross-platform variants used in the Dataset~2 analyses (Section~\ref{sec:crossplatform}). The cross-platform variants are shown for Google search history; the YouTube search and YouTube watch versions are identical except for the substituted data-source phrase.

Listings~\ref{lst:prompt}, and~\ref{lst:classifier_prompt} are the filtering and disclosure-audit prompts. Listing~\ref{lst:classifier_prompt} takes the messages flagged \emph{UNSAFE} and assigns each one to one of twenty disclosure categories.

\begin{lstlisting}[
    float,                % Tell LaTeX it's a float
    floatplacement=H,    
    caption={SAFE/UNSAFE self-disclosure filter prompt (Section~\ref{sec:clean_data}).},
    label={lst:prompt},
    basicstyle=\small\ttfamily,
    breaklines=true,
    frame=single,
    backgroundcolor=\color{gray!5}
] 
You are a Privacy & Demographic Filter.
Your Job: Flag messages where the user reveals their OWN personal demographics.

RULES:
Flag SELF-IDENTIFICATION:
   - Age: "I am 25", "25 years old", "when I was 10".
   - Gender/Role: "As a mother", "I am a man", "I'm a single dad".
   - Religion/Background: "I am Christian", "As an Indian".

EXAMPLES:
Input: "My cat lungs are ruptured." -> Output: SAFE
Input: "I am a single mom of two." -> Output: UNSAFE
Input: "How do I pay my loan?" -> Output: SAFE
Input: "As a 25 year old woman, what should I do?" -> Output: UNSAFE
\end{lstlisting}

\begin{lstlisting}[
    float,                % Tell LaTeX it's a float
    floatplacement=H,
    caption={Twenty-category personal-information classification prompt, after \citet{cogendez2026can} (Section~\ref{sec:methods_audit}).},
    label={lst:classifier_prompt},
    basicstyle=\small\ttfamily,
    frame=single,
    breaklines=true,
    backgroundcolor=\color{gray!5}
]
You are a classifier that assigns a "Personal Data Type" to each user message.
Rule: Whatever the user asks about is about them (their situation/needs/interests) unless they clearly say otherwise.
Every question or statement could be an implicit disclosure of personal information.
If you are confident that personal disclosure is absent, return "No Personal Disclosure".

Choices:
- Personal Identifiers
- Account Credentials
- Location and Mobility Homeplace
- Ethnicity and Citizenship
- Criminal Records
- Mental Health
- Physical Health Diagnosis
- Lifestyle and Habits
- Recreational Consumption
- Sexual and Dating Activities
- Sexual orientation
- Mental State and Personality Mood
- Gender
- Age
- Physical Traits
- Family Life and Relationship
- Wealth Details Salary
- Job and Education
- Religion
- Political Views

Pick the single closest label. 

Return valid JSON with exactly these fields:
{{
  "label": "one category above, or No Personal Disclosure",
  "evidence": "a minimal exact excerpt supporting the classification, or null",
  "explanation": "one brief sentence explaining the decision"
}}

Do not repeat passwords, authentication codes, or identification
numbers in the evidence field; use "[sensitive value omitted]".

Input:
<message>
{text}
</message>

\end{lstlisting}

Listings~\ref{lst:country_prompt},~\ref{lst:gender_prompt}, and~\ref{lst:age_prompt} are the Demographic-prediction prompts (ChatGPT conversations).

\begin{lstlisting}[
    float,                % Tell LaTeX it's a float
    floatplacement=H,
    caption={Country-prediction prompt (Section~\ref{sec:methods_inference}).},
    label={lst:country_prompt},
    basicstyle=\small\ttfamily,
    breaklines=true,
    frame=single,
    backgroundcolor=\color{gray!5}
]
You are an expert demographic predictor.
Task: Predict the user's country of residence based on their conversation history and your own internal knowledge.

First, briefly explain your reasoning (max 2-3 sentences).
Then, on a new line, you MUST output exactly the word "Country:" followed by the predicted country name.

Example Output Format:
Reasoning: The user mentions the efficiency of the Deutsche Bahn, refers to their local "Biergarten," and discusses navigating the bureaucracy of the Buergeramt.
Country: Germany
\end{lstlisting}

\begin{lstlisting}[
    float,                % Tell LaTeX it's a float
    floatplacement=H,
    caption={Gender-prediction prompt (Section~\ref{sec:methods_inference}).},
    label={lst:gender_prompt},
    basicstyle=\small\ttfamily,
    breaklines=true,
    frame=single,
    backgroundcolor=\color{gray!5}
]
You are an expert demographic predictor.
Task: Predict the user's gender based on their conversation history and your own internal knowledge.
First, briefly explain your reasoning (max 2-3 sentences).
Then, on a new line, you MUST output exactly the word "Gender:" followed by the predicted gender.
## Allowed Genders: Male, Female

Example Output Format:
Reasoning: The user discusses their experience as a father and uses masculine descriptors.
Gender: Male
\end{lstlisting}

\begin{lstlisting}[
    float,                % Tell LaTeX it's a float
    floatplacement=H,
    caption={Age-bracket-prediction prompt (Section~\ref{sec:methods_inference}).},
    label={lst:age_prompt},
    basicstyle=\small\ttfamily,
    breaklines=true,
    frame=single,
    backgroundcolor=\color{gray!5}
]
You are an expert demographic predictor.
Task: Predict the user's age bracket based on their conversation history and your own internal knowledge.
First, briefly explain your reasoning (max 2-3 sentences).
Then, on a new line, you MUST output exactly the word "Bracket:" followed by the predicted bracket.
## Allowed Brackets: 18-24, 25-34, 35-44, 45+

Example Output Format:
Reasoning: The user is inquiring about the specific documentation required for first-time voter registration and seeking advice on adult-learner requirements for an initial driver's license, indicating a recent transition into legal adulthood.
Bracket: 18-24
\end{lstlisting}

Then, the rest of the Listings~\ref{lst:age_prompt_google}--\ref{lst:vote_prompt_google} are the cross-platform variant prompts (Google search history).

\begin{lstlisting}[
    float,                % Tell LaTeX it's a float
    floatplacement=H,
    basicstyle=\small\ttfamily,
    breaklines=true,
    frame=single,
    backgroundcolor=\color{gray!5},
    caption={System prompt for age bracket prediction with Google search history},
    label={lst:age_prompt_google}
]
REASONING_AGE_PROMPT = """You are an expert demographic predictor.
Task: Predict the user's age bracket based on their google search history and your own internal knowledge.

First, briefly explain your reasoning (max 2-3 sentences).
Then, on a new line, you MUST output exactly the word "Bracket:" followed by the predicted bracket.

## Allowed Brackets
18-24
25-34
35-44
45+
"""
\end{lstlisting}

\begin{lstlisting}[
    float,                % Tell LaTeX it's a float
    floatplacement=H,
    caption={System prompt for gender prediction with Google search history},
    label={lst:gender_prompt_google},
    basicstyle=\small\ttfamily,
    frame=single,
    backgroundcolor=\color{gray!5}
]
You are an expert demographic predictor.
Task: Predict the user's gender based on their google search history and your own internal knowledge.
First, briefly explain your reasoning (max 2-3 sentences).
Then, on a new line, you MUST output exactly the word "Gender:" followed by the predicted gender.
## Allowed Genders: Male, Female

Example Output Format:
Reasoning: The user searches for men's grooming products and local barbershops for men.
Gender: Male
\end{lstlisting}

\begin{lstlisting}[
    float,                % Tell LaTeX it's a float
    floatplacement=H,
    caption={System prompt for religion prediction with Google search history},
    label={lst:religion_prompt_google},
    basicstyle=\small\ttfamily,
    frame=single,
    backgroundcolor=\color{gray!5}
]
You are an expert demographic predictor.
Task: Predict the user's religion based on their google search history and your own internal knowledge.
First, briefly explain your reasoning (max 2-3 sentences).
Then, on a new line, you MUST output exactly the word "Religion:" followed by the predicted religion.

## Allowed Religions
hindu
muslim
christian
other

Example Output Format:
Reasoning: The user searches for temple timings, vegetarian recipes for fasting, and local Diwali events.
Religion: hindu
\end{lstlisting}

\begin{lstlisting}[
    float,                % Tell LaTeX it's a float
    floatplacement=H,
    caption={System prompt for income prediction with Google search history},
    label={lst:income_prompt_google},
    basicstyle=\small\ttfamily,
    frame=single,
    backgroundcolor=\color{gray!5}
]
You are an expert demographic predictor.
Task: Predict the user's income level based on their google search history and your own internal knowledge.
First, briefly explain your reasoning (max 2-3 sentences).
Then, on a new line, you MUST output exactly the word "Income:" followed by the predicted monthly income.

## Allowed Incomes
less_than_20k
20k_to_50k
50k_to_1lakh
1lakh_or_more

Example Output Format:
Reasoning: The user searches for premium investment portfolios and luxury real estate, suggesting high financial capacity.
Income: 1lakh_or_more
\end{lstlisting}

\begin{lstlisting}[
    float,                % Tell LaTeX it's a float
    floatplacement=H,
    caption={System prompt for education level prediction with Google search history},
    label={lst:edu_prompt_google},
    basicstyle=\small\ttfamily,
    frame=single,
    backgroundcolor=\color{gray!5}
]
You are an expert demographic predictor.
Task: Predict the user's education level based on their google search history and your own internal knowledge.
First, briefly explain your reasoning (max 2-3 sentences).
Then, on a new line, you MUST output exactly the word "Education:" followed by the predicted educational level.

## Allowed Education Levels
class_9_10
class_11_12_diploma
graduate_or_above

Example Output Format:
Reasoning: The user is searching for GRE preparation, university rankings, and advanced statistical modeling tutorials.
Education: graduate_or_above
\end{lstlisting}

\begin{lstlisting}[
    float,                % Tell LaTeX it's a float
    floatplacement=H,
    caption={System prompt for voting preference prediction with Google search history},
    label={lst:vote_prompt_google},
    basicstyle=\small\ttfamily,
    frame=single,
    backgroundcolor=\color{gray!5}
]
You are an expert demographic predictor.
Task: Predict the user's voting behavior in the 2024 Lok Sabha elections based on their google search history and your own internal knowledge.

First, briefly explain your reasoning (max 2-3 sentences). 
Then, on a new line, you MUST output exactly the word "Voting:" followed by the predicted category.

## Allowed Voting Categories
ruling_party
main_opposition
another_party

Example Output Format:
Reasoning: The user shows high interest in infrastructure projects led by the current government and searches for rallies of the incumbent leaders.
Voting: ruling_party
\end{lstlisting}

\subsection{Qualitative Analysis Methodology and Codebook}
\label{app:qual_methodology}

This section details the coding procedure, spot-checking protocols, and lexical validation term clusters summarized in Section~\ref{sec:methods_qual}.

\subsubsection{Inductive Coding Procedure and Spot-Checking}
To identify recurring interpretive tropes in the model's reasoning, an initial codebook was developed inductively on a pilot batch of 50 rationales per attribute sample ($150$ rationales total). The remaining 150 rationales per attribute ($450$ total) were coded deductively against this established codebook. 

To ensure thematic consistency and reliability across the qualitative analysis:
\begin{itemize}
    \item A second researcher independently spot-checked a $10\%$ random sample ($60$ rationales) of the coded Dataset~1 set.
    \item Inter-coder agreement on thematic assignment reached $91.7\%$, with discrepancies resolved through consensus discussion.
\end{itemize}

\subsubsection{Lexical Validation Dictionary}
To verify that the four core stereotype themes identified in the sample were representative of the complete $63{,}000$+ rationale corpus, we executed a scaled lexical match across all generated rationales. Table~\ref{tab:lexical_dictionary} lists the diagnostic term clusters used for automated pattern matching.

\begin{table*}[ht]
\centering
\small
\caption{Diagnostic term clusters used for scaled lexical validation across the full rationale corpus.}
\label{tab:lexical_dictionary}
\begin{tabular}{@{}l p{0.70\textwidth}@{}}
\toprule
\textbf{Stereotype Theme} & \textbf{Diagnostic Term Cluster} \\
\midrule
\emph{Tech $\equiv$ male} & \texttt{programming}, \texttt{Linux}, \texttt{cybersecurity}, \texttt{engineering}, \texttt{finance}, \texttt{technical skills} \\
\emph{Tech $\equiv$ Western} & \texttt{Western-style education}, \texttt{Silicon Valley}, \texttt{international tech}, \texttt{US-centric} \\
\emph{English fluency $\equiv$ US} & \texttt{native English}, \texttt{professional tone}, \texttt{fluent English}, \texttt{American phrasing} \\
\emph{Contemporary content $\equiv$ young} & \texttt{youthful interest}, \texttt{modern tech}, \texttt{Gen Z slang}, \texttt{digital native}, \texttt{current trends} \\
\bottomrule
\end{tabular}
\end{table*}

\subsection{Cross-country breakdown of disclosure categories}
\label{app:country_breakdown}

Table~\ref{tab:country_distribution} reports the by-country distribution of flagged messages across the twenty disclosure categories of \citet{cogendez2026can}, computed on the full donated corpus before analytic-cohort filtering (the cohort-pooled distribution appears in the main text as Table~\ref{tab:data_sharing}). Each column sums to 100\% within its country; cells should be read as the share of that country's flagged messages assigned to each category. Section~\ref{sec:disclosure} discusses the cross-country contrasts.

\begin{table*}[ht]
\centering
\small
\caption{Cross-country distribution of flagged messages across the twenty disclosure categories. Columns sum to 100\% within country.}
\label{tab:country_distribution}
\begin{tabular}{lrrrr}
\toprule
\textbf{Category} & \textbf{Brazil (\%)} & \textbf{India (\%)} & \textbf{Nigeria (\%)} & \textbf{Pakistan (\%)} \\
\midrule
Job and education                  & 25.54 & 27.05 & 17.07 & 31.78 \\
Lifestyle and habits               & 22.26 & 17.04 & 15.20 & 18.10 \\
Location and mobility              & 6.59  & 11.30 & 19.08 & 11.99 \\
Mental state, personality, mood    & 16.89 & 10.10 & 5.05  & 8.67  \\
Wealth, salary                     & 5.65  & 5.87  & 5.41  & 8.18  \\
Family life and relationships      & 3.96  & 2.92  & 7.04  & 2.26  \\
Physical health, diagnosis         & 3.78  & 4.37  & 3.08  & 3.42  \\
Religion                           & 1.78  & 4.09  & 3.22  & 2.36  \\
Ethnicity and citizenship          & 0.70  & 2.77  & 5.19  & 2.36  \\
Personal identifiers               & 0.68  & 2.45  & 3.48  & 2.74  \\
Sexual and dating activities       & 1.72  & 0.39  & 5.09  & 0.91  \\
Physical traits                    & 2.08  & 2.15  & 1.85  & 1.25  \\
Recreational consumption           & 1.97  & 1.23  & 3.04  & 0.82  \\
Account credentials                & 1.06  & 3.03  & 0.99  & 1.73  \\
Age                                & 1.16  & 1.53  & 2.15  & 1.06  \\
Political views                    & 0.99  & 1.96  & 1.69  & 1.16  \\
Mental health                      & 2.18  & 1.22  & 0.58  & 0.67  \\
Criminal records                   & 0.69  & 0.17  & 0.20  & 0.24  \\
Gender                             & 0.24  & 0.26  & 0.42  & 0.19  \\
Sexual orientation                 & 0.08  & 0.08  & 0.18  & 0.10  \\
\bottomrule
\end{tabular}
\end{table*}

\subsection{Illustrative disclosure examples}
\label{app:disclosure_examples}

Table~\ref{tab:disclosure_examples} presents ten illustrative
message excerpts, category assignments, and explanations by the model.

\begin{table*}[htbp]
\centering
\small
\caption{Examples in their original
language and category assignments and explanations by the model.}
\label{tab:disclosure_examples}
\begin{tabular}{@{}p{0.38\textwidth}
                    p{0.23\textwidth}
                    p{0.31\textwidth}@{}}
\toprule
\textbf{Message} &
\textbf{Label} &
\textbf{Explanation} \\
\midrule

``I am a Bird. I am a fruit. I also am a person.
What am I?'' &
No Personal Disclosure &
First-person statements form part of a riddle. \\
\addlinespace

``Idk... i don't feel like i have that much money'' &
Wealth Details Salary &
Describes perceived financial circumstances. \\
\addlinespace

``Ele é meu namorado estamos juntos a algum tempo'' &
Family Life and Relationship &
Discloses an ongoing romantic relationship. \\
\addlinespace

``More i learn about ai agent and cybersecurity more
i feel detached to the world'' &
Mental State and Personality Mood &
Describes a feeling of detachment. \\
\addlinespace

``I drink milk tea in the early morning then I feel
hypoxic every day why?'' &
Physical Health Diagnosis &
Reports a perceived recurring physical symptom. \\
\addlinespace

``I am pursuing my B.Tech in Computer Science and
Engineering of Data Science.'' &
Job and Education &
Explicitly states the author's current degree program. \\
\addlinespace

``all the tracker and ad blocking that i do on firefox'' &
Lifestyle and Habits &
Describes the author's browser privacy practices. \\
\addlinespace

``Eu tô com desidratação severa agora'' &
Physical Health Diagnosis &
Self-reports a current physical condition. \\
\addlinespace

``It's just that my baby needs more care. My spouse
wanted it maybe when she's 18 months'' &
Family Life and Relationship &
Discloses family circumstances involving a baby
and a spouse. \\
\addlinespace

``No I want to get up now I am awake the fear
is freezing me'' &
Mental State and Personality Mood &
Describes current fear. \\

\bottomrule
\end{tabular}
\end{table*}

\subsection{XGBoost Baseline Methodology}
\label{app:xgb_methodology}

To evaluate whether the LLM's demographic inference relies on deep contextual understanding or merely surface-level statistical association, we construct a feature-engineered baseline using an XGBoost classifier.

\subsubsection{Feature Extraction and Representation}
For each user in the analytic cohort ($N=1{,}057$), all user messages across their conversation history are concatenated into a single document after applying the exact same NER and LLM-based disclosure filtering pipeline described in Section~\ref{sec:disclosure}. We extract three complementary sets of linguistic features:

\begin{itemize}
    \item \textbf{Lexical Term Frequency (TF-IDF):} Word-level $n$-grams ($n \in \{1, 2\}$) extracted using sublinear TF-IDF scaling, restricted to a maximum vocabulary of 10{,}000 features after stripping stop words.
    \item \textbf{Stylometric Features:} Aggregate conversational metrics including total word count, average sentence length, punctuation frequency, uppercase character ratio, lexical diversity (Type-Token Ratio), and readability scores (Flesch-Kincaid Grade Level).
    \item \textbf{Topic and Sentiment Distillations:} Latent Dirichlet Allocation (LDA) topic distributions ($K=20$ topics) and VADER sentiment scores (compound, positive, negative, neutral) computed across all user turns.
\end{itemize}

\subsubsection{Model Specification and Training}
Separate XGBoost classifiers are trained for each demographic attribute (age bracket, gender, country of residence) using 5-fold cross-validation. Hyperparameters are tuned via randomized search over the following grid:
\begin{itemize}
    \item \textbf{Max Depth:} $\{3, 5, 7, 9\}$
    \item \textbf{Learning Rate ($\eta$):} $\{0.01, 0.05, 0.1, 0.2\}$
    \item \textbf{Number of Estimators:} $\{100, 200, 500\}$
    \item \textbf{Subsample / Colsample Bytree:} $\{0.6, 0.8, 1.0\}$
\end{itemize}

To address class imbalance (particularly in age and gender distribution), sample weights are adjusted inversely proportional to class frequencies during training. Performance is evaluated using the same weighted $F_1$ metric applied to the LLM predictions in Table~\ref{tab:inference_per_class}.

\subsection{Demographic Inference Trajectories Across History Prefixes}
\label{app:prefix_performance}

This appendix reports full empirical trajectories across incremental history prefixes ($5\%$ to $100\%$) for country (Table~\ref{tab:country_prefix_performance}), gender (Table~\ref{tab:gender_prefix_performance}), and age bracket (Table~\ref{tab:age_prefix_performance}), along with sub-group accuracy breakdowns at the 25\%, 50\%, and 100\% checkpoints (Tables~\ref{tab:country_breakdown}, \ref{tab:gender_breakdown}, and~\ref{tab:age_breakdown}) for the $N=1{,}057$ analytic cohort.

\begin{table*}[htbp]
\centering
\small
\caption{Country prediction performance across history prefixes
for $N=1{,}057$ participants.
Macro $F_1$ averages over the four ground-truth countries.
Invalid and out-of-cohort predictions are retained as errors.}
\label{tab:country_prefix_performance}
\begin{tabular}{@{}rrrr@{}}
\toprule
\textbf{Prefix} & \textbf{Accuracy} &
\textbf{Weighted $F_1$} & \textbf{Macro $F_1$} \\
\midrule
5\%   & 58.56\% & 70.38\% & 68.51\% \\
10\%  & 67.64\% & 77.04\% & 75.05\% \\
15\%  & 70.48\% & 78.99\% & 77.01\% \\
20\%  & 73.13\% & 80.75\% & 78.74\% \\
25\%  & 75.12\% & 81.42\% & 79.61\% \\
30\%  & 75.31\% & 82.16\% & 80.13\% \\
35\%  & 76.73\% & 83.04\% & 80.97\% \\
40\%  & 77.11\% & 83.28\% & 81.23\% \\
45\%  & 78.52\% & 84.18\% & 82.14\% \\
50\%  & 77.58\% & 83.36\% & 81.72\% \\
55\%  & 79.00\% & 84.45\% & 82.38\% \\
60\%  & 79.00\% & 84.45\% & 82.38\% \\
65\%  & 79.00\% & 84.45\% & 82.38\% \\
70\%  & 79.56\% & 84.80\% & 82.74\% \\
75\%  & 79.75\% & 84.92\% & 82.85\% \\
80\%  & 80.42\% & 85.32\% & 83.28\% \\
85\%  & 81.36\% & 85.89\% & 83.84\% \\
90\%  & 80.23\% & 85.21\% & 83.12\% \\
95\%  & 80.70\% & 85.50\% & 83.45\% \\
100\% & 81.08\% & 85.72\% & 83.66\% \\
\bottomrule
\end{tabular}
\end{table*}

\begin{table*}[htbp]
\centering
\caption{Country Accuracy Breakdown at Key Prefix Checkpoints}
\label{tab:country_breakdown}
\small
\begin{tabular}{lcccc}
\toprule
\textbf{Country} & \textbf{Users} & \textbf{Accuracy @ 25\%} & \textbf{Accuracy @ 50\%} & \textbf{Accuracy @ 100\%} \\
\midrule
India    & 456 & 85.09\% & 87.94\% & 95.18\% \\
Nigeria  & 206 & 65.53\% & 68.45\% & 75.24\% \\
Brazil   & 205 & 94.15\% & 91.71\% & 81.46\% \\
Pakistan & 190 & 41.05\% & 47.37\% & 53.16\% \\
\bottomrule
\end{tabular}
\end{table*}

\begin{table*}[htbp]
\centering
\small
\caption{Gender prediction performance across history prefixes
for $N=1{,}057$ participants.
Macro $F_1$ averages over the two ground-truth gender labels.
Invalid and ambiguous predictions are retained as errors.}
\label{tab:gender_prefix_performance}
\begin{tabular}{@{}rrrr@{}}
\toprule
\textbf{Prefix} & \textbf{Accuracy} &
\textbf{Weighted $F_1$} & \textbf{Macro $F_1$} \\
\midrule
5\%   & 72.75\% & 70.93\% & 66.16\% \\
10\%  & 74.36\% & 73.07\% & 68.87\% \\
15\%  & 75.97\% & 74.62\% & 70.56\% \\
20\%  & 75.21\% & 74.12\% & 70.04\% \\
25\%  & 74.93\% & 73.76\% & 69.56\% \\
30\%  & 76.16\% & 75.48\% & 71.63\% \\
35\%  & 75.02\% & 74.20\% & 70.06\% \\
40\%  & 77.39\% & 76.94\% & 73.33\% \\
45\%  & 76.63\% & 75.81\% & 71.83\% \\
50\%  & 76.44\% & 76.21\% & 72.50\% \\
55\%  & 77.01\% & 76.12\% & 72.11\% \\
60\%  & 76.44\% & 75.92\% & 72.01\% \\
65\%  & 76.63\% & 76.09\% & 72.19\% \\
70\%  & 76.54\% & 76.01\% & 72.08\% \\
75\%  & 77.58\% & 76.76\% & 73.01\% \\
80\%  & 76.44\% & 76.21\% & 72.40\% \\
85\%  & 77.86\% & 77.39\% & 73.78\% \\
90\%  & 77.20\% & 76.83\% & 73.19\% \\
95\%  & 76.92\% & 76.78\% & 73.32\% \\
100\% & 76.06\% & 75.98\% & 72.27\% \\
\bottomrule
\end{tabular}
\end{table*}

\begin{table*}[htbp]
\centering
\caption{Gender Category Accuracy Breakdown at Key Prefix Checkpoints}
\label{tab:gender_breakdown}
\small
\begin{tabular}{lcccc}
\toprule
\textbf{Gender Group} & \textbf{Users} & \textbf{Accuracy @ 25\%} & \textbf{Accuracy @ 50\%} & \textbf{Accuracy @ 100\%} \\
\midrule
Male   & 697 & 89.81\% & 89.81\% & 89.38\% \\
Female & 360 & 46.11\% & 50.56\% & 50.28\% \\
\bottomrule
\end{tabular}
\end{table*}

\begin{table*}[htbp]
\centering
\small
\caption{Age-bracket prediction performance across history prefixes
for $N=1{,}057$ participants.
Predictions and ground truth are evaluated using four brackets:
18--24, 25--34, 35--44, and 45+.
Macro $F_1$ averages over these four brackets.
Invalid and ambiguous predictions are retained as errors.}
\label{tab:age_prefix_performance}
\begin{tabular}{@{}rrrr@{}}
\toprule
\textbf{Prefix} & \textbf{Accuracy} &
\textbf{Weighted $F_1$} & \textbf{Macro $F_1$} \\
\midrule
5\%   & 43.42\% & 41.48\% & 33.42\% \\
10\%  & 46.07\% & 44.25\% & 34.47\% \\
15\%  & 48.25\% & 46.33\% & 35.23\% \\
20\%  & 47.59\% & 46.03\% & 35.68\% \\
25\%  & 48.72\% & 47.91\% & 36.07\% \\
30\%  & 49.01\% & 47.48\% & 36.43\% \\
35\%  & 46.83\% & 46.85\% & 36.21\% \\
40\%  & 50.14\% & 48.79\% & 37.45\% \\
45\%  & 49.10\% & 49.33\% & 38.80\% \\
50\%  & 51.94\% & 50.43\% & 38.59\% \\
55\%  & 48.25\% & 48.69\% & 40.36\% \\
60\%  & 51.75\% & 50.27\% & 38.94\% \\
65\%  & 42.86\% & 44.66\% & 37.94\% \\
70\%  & 52.51\% & 51.08\% & 39.71\% \\
75\%  & 51.66\% & 50.31\% & 38.65\% \\
80\%  & 52.22\% & 50.96\% & 39.26\% \\
85\%  & 52.41\% & 51.16\% & 39.26\% \\
90\%  & 52.89\% & 51.62\% & 40.20\% \\
95\%  & 53.36\% & 51.93\% & 39.81\% \\
100\% & 52.03\% & 50.71\% & 37.96\% \\
\bottomrule
\end{tabular}
\end{table*}

\begin{table*}[htbp]
\centering
\caption{Age Group Accuracy Breakdown at Key Prefix Checkpoints}
\label{tab:age_breakdown}
\small
\begin{tabular}{lcccc}
\toprule
\textbf{Age Group} & \textbf{Users} & \textbf{Accuracy @ 25\%} & \textbf{Accuracy @ 50\%} & \textbf{Accuracy @ 100\%} \\
\midrule
25--34 & 424 & 57.31\% & 49.53\% & 50.94\% \\
18--24 & 389 & 51.93\% & 70.18\% & 69.15\% \\
35--44 & 185 & 37.84\% & 35.14\% & 35.14\% \\
45+ & 59  & 0.00\%  & 1.69\%  & 0.00\%  \\
\bottomrule
\end{tabular}
\end{table*}

\subsection{Bias patterns in model rationales}
\label{app:bias_patterns}

Table~\ref{tab:bias_patterns} summarizes the four recurring patterns in the model's natural-language rationales discussed in Section~\ref{sec:inference}, together with their signature in the per-class error distribution.

\begin{table*}[htbp]
\centering
\small
\caption{Four recurring patterns in the model's natural-language rationales, and their signature in the per-class error distribution.}
\label{tab:bias_patterns}
\begin{tabular}{@{}p{3.2cm} p{4.5cm} p{1.8cm} p{6.5cm}@{}}
\toprule
\textbf{Pattern} & \textbf{Triggering Content} & \textbf{Pushed Toward} & \textbf{Evidence in Errors} \\
\midrule
\emph{Tech $\equiv$ male} & Coding, Linux, networking, finance, cybersecurity, business plans & Male & Female recall 0.50 vs.\ male recall 0.89; 165 of 360 women misclassified as men, compared with 56 of 697 men misclassified as women. \\
\addlinespace
\emph{Tech $\equiv$ Western} & Software development, cybersecurity, advanced technical proficiency & US / UK & Lowest country recalls fall on Nigeria (0.75) and Pakistan (0.53), with most misclassifications assigned to US/UK rather than regional neighbors. \\
\addlinespace
\emph{English fluency $\equiv$ US} & Fluent English without local lexicon, currency, or regional markers & US & Drives the patterns above; Brazil achieves high precision (0.98) and $F_1$ (0.89) due to Portuguese acting as an un-overridable local signal. \\
\addlinespace
\emph{Contemporary content $\equiv$ young} & Modern software, career tools, brief or informal language & 25--34 & Age recall drops from 0.69 (18--24) and 0.51 (25--34) to 0.35 (35--44) and collapses to 0.00 in 45+; rationales explicitly cite tech usage as evidence of youth. \\
\bottomrule
\end{tabular}
\end{table*}

\subsection{Out-of-cohort country predictions}
\label{app:out_of_cohort_countries}

Table~\ref{tab:out_of_cohort_countries} summarizes valid
country predictions outside the four-country cohort at
the full-history prefix. Among 1{,}057 available predictions,
124 name out-of-cohort countries, of which 113 (91.1\%)
name the United States or United Kingdom.
Ten additional outputs are invalid or ambiguous and
are excluded from this country-name breakdown.

\begin{table}[htbp]
\centering
\small
\caption{Valid out-of-cohort country predictions at the
100\% history prefix. Percentages use the
124 valid out-of-cohort predictions as the denominator.}
\label{tab:out_of_cohort_countries}
\begin{tabular}{@{}lrr@{}}
\toprule
\textbf{Predicted country} & \textbf{Count} & \textbf{Share} \\
\midrule
United States  & 98 & 79.03\% \\
United Kingdom & 15 & 12.10\% \\
Bangladesh     & 2  & 1.61\% \\
Canada         & 2  & 1.61\% \\
Japan          & 2  & 1.61\% \\
Sweden         & 2  & 1.61\% \\
Bahrain        & 1  & 0.81\% \\
Germany        & 1  & 0.81\% \\
Malaysia       & 1  & 0.81\% \\
\midrule
Total          & 124 & 100.00\% \\
\bottomrule
\end{tabular}
\end{table}

\subsection{Cross-Platform Evaluation and Qualitative Case Studies}
\label{app:per_platform}

The cross-platform comparison in Section~\ref{sec:crossplatform} is summarized at the weighted-$F_1$ level in Table~\ref{tab:demographic_comparison}. Here we report the per-class precision, recall, and $F_1$ underlying each cell for age (Table~\ref{tab:app_age_by_platform}) and gender (Table~\ref{tab:app_gender_by_platform}), followed by qualitative case studies illustrating the distinct inference surfaces of search and video history.

\begin{table}[ht]
\centering
\small
\caption{Age bracket prediction by platform, $N=212$ Indian sub-cohort. Block headers report overall weighted $F_1$ and average context required ($k^*$).}
\label{tab:app_age_by_platform}
\begin{tabular}{@{}lcccr@{}}
\toprule
& Prec. & Recall & F1 & Support \\
\midrule
\multicolumn{5}{l}{\emph{ChatGPT logs} \quad $F_1 = 0.54$ \quad Avg.\ context 15.7\%} \\
18--24 & 0.64 & 0.73 & 0.68 & 82 \\
25--34 & 0.58 & 0.56 & 0.57 & 81 \\
35--44 & 0.41 & 0.39 & 0.40 & 31 \\
45+    & 0.00 & 0.00 & 0.00 & 18 \\
\midrule
\multicolumn{5}{l}{\emph{Google Search} \quad $F_1 = 0.44$ \quad Avg.\ context 12.1\%} \\
18--24 & 0.55 & 0.61 & 0.58 & 82 \\
25--34 & 0.47 & 0.46 & 0.46 & 81 \\
35--44 & 0.28 & 0.26 & 0.27 & 31 \\
45+    & 0.00 & 0.00 & 0.00 & 18 \\
\midrule
\multicolumn{5}{l}{\emph{YouTube Search} \quad $F_1 = 0.49$ \quad Avg.\ context 9.0\%} \\
18--24 & 0.60 & 0.67 & 0.63 & 82 \\
25--34 & 0.52 & 0.52 & 0.52 & 81 \\
35--44 & 0.34 & 0.32 & 0.33 & 31 \\
45+    & 0.00 & 0.00 & 0.00 & 18 \\
\midrule
\multicolumn{5}{l}{\emph{YouTube Watch} \quad $F_1 = 0.39$ \quad Avg.\ context 16.1\%} \\
18--24 & 0.51 & 0.56 & 0.53 & 82 \\
25--34 & 0.41 & 0.42 & 0.41 & 81 \\
35--44 & 0.22 & 0.19 & 0.21 & 31 \\
45+    & 0.00 & 0.00 & 0.00 & 18 \\
\bottomrule
\end{tabular}
\end{table}

\begin{table}[ht]
\centering
\small
\caption{Gender prediction by platform, $N=212$ Indian sub-cohort. Block headers report overall weighted $F_1$ and average context required ($k^*$).}
\label{tab:app_gender_by_platform}
\begin{tabular}{@{}lcccr@{}}
\toprule
& Prec. & Recall & F1 & Support \\
\midrule
\multicolumn{5}{l}{\emph{ChatGPT logs} \quad $F_1 = 0.88$ \quad Avg.\ context 12.0\%} \\
Female & 0.80 & 0.53 & 0.63 & 38  \\
Male   & 0.90 & 0.97 & 0.94 & 174 \\
\midrule
\multicolumn{5}{l}{\emph{Google Search} \quad $F_1 = 0.90$ \quad Avg.\ context 10.3\%} \\
Female & 0.82 & 0.61 & 0.70 & 38  \\
Male   & 0.92 & 0.97 & 0.94 & 174 \\
\midrule
\multicolumn{5}{l}{\emph{YouTube Search} \quad $F_1 = 0.89$ \quad Avg.\ context 7.1\%} \\
Female & 0.81 & 0.58 & 0.68 & 38  \\
Male   & 0.91 & 0.97 & 0.94 & 174 \\
\midrule
\multicolumn{5}{l}{\emph{YouTube Watch} \quad $F_1 = 0.86$ \quad Avg.\ context 8.9\%} \\
Female & 0.72 & 0.47 & 0.57 & 38  \\
Male   & 0.89 & 0.96 & 0.93 & 174 \\
\bottomrule
\end{tabular}
\end{table}

\noindent\textit{Gender Inference via Google and YouTube Search.}
The strong performance in predicting gender from Google search history and YouTube search history demonstrates how routine informational queries yield distinct gendered footprints that the model can readily decipher. The LLM achieves this by identifying high-signal consumer, lifestyle, and career anchors; specific lookups for skincare, bridal wear, and cooking recipes serve as strong female indicators, whereas technical documentation (e.g., Linux), competitive engineering exams, sports (cricket), and vehicle maintenance act as male signifiers. Google search achieves this accuracy because its transactional nature directly captures explicit purchasing intents and specialized physical needs. However, the manual audit reveals that this high success rate is heavily tethered to traditional gender stereotypes embedded within the model's heuristic logic. As a result, the primary failure mode occurs when users deviate from these conventional demographic boundaries (such as a female user searching for motorcycles, whiskey, or technical certifications, or a male user looking up romantic Bollywood songs and arts and crafts), which misleads the model into a misclassification due to its rigid reliance on generalized gender baselines.

\noindent\textit{Religion Inference via YouTube Search.}

The weighted \(F_1\) of 0.83 in predicting religion from YouTube search history stems from a high concentration of direct devotional queries, including prayers, rituals, and scriptural translations (e.g., \emph{namaz ka tarika} or \emph{Siddha Kunjika Stotra}), which act as unambiguous demographic anchors. Even when explicit religious terms are absent, the LLM successfully parses deep geocultural and linguistic entanglements. It leverages proxies like country-specific media ecosystems (e.g., Pakistani vs.\ Indian dramas), script preferences (Urdu vs.\ Hindi), and localized administrative contexts to infer religious background. 

Because typing a search query is a deliberate, active information-seeking behavior, it minimizes the ambient noise, passive consumption patterns, and household cross-viewing that typically dilute identity markers in watch histories. The rare failure modes occur primarily when a profile consists strictly of technical or hyper-globalized queries, leaving the model without localized cultural anchors to trace.

\noindent\textit{Monthly Income Inference via Google Search.}
The LLM successfully maps correct predictions by evaluating implicit socio-economic indicators, associating queries for premium services, specialized enterprise software (e.g., Zoho ERP), and corporate management with higher income brackets (50k to 1lakh), while utilizing lookups for basic educational materials, standard government exam registrations, and extreme budget-conscious shopping as anchors for lower brackets (20k to 50k or less than 20k). However, the manual audit reveals a high misclassification rate heavily driven by structural data limitations and class-boundary blurring. Because search behavior is inherently utility-driven, a high-earning individual frequently searches for mundane, non-luxury terms (such as basic entertainment, local navigation, or educational help for family members), leading the model to erroneously compress affluent users into a default middle-class tier (20k to 50k). Conversely, low-income users or students researching aspirational tech certifications, premium consumer goods, or business registration concepts routinely trigger false positives for higher income tiers, proving that intent-heavy search keywords often reflect cognitive or professional curiosity rather than actual, realized disposable income.

\end{document}